\documentclass[journal]{IEEEtran}
\usepackage{ifpdf}
 \ifpdf
 \else
 \fi
\usepackage{color}
\usepackage{cite}
\ifCLASSINFOpdf
   \usepackage[pdftex]{graphicx}
   \usepackage{epstopdf}
  \graphicspath{{fig/}}
  \DeclareGraphicsExtensions{.eps,.pdf,.jpg,.png,.svg}
\else
\fi
\usepackage{amsmath}
\usepackage{mathrsfs}
\usepackage{amssymb}
\usepackage{booktabs}
\usepackage{multirow}

\usepackage{algorithm}
\usepackage{algpseudocode}

\usepackage{amsmath,amssymb,amsfonts}
\usepackage{array}
\usepackage{bm}

\ifCLASSOPTIONcompsoc
  \usepackage[caption=false,font=normalsize,labelfont=sf,textfont=sf]{subfig}
\else
  \usepackage[caption=false,font=footnotesize]{subfig}
\fi

\usepackage{stfloats}
\begin{document}
%
% paper title
% Titles are generally capitalized except for words such as a, an, and, as,
% at, but, by, for, in, nor, of, on, or, the, to and up, which are usually
% not capitalized unless they are the first or last word of the title.
% Linebreaks \\ can be used within to get better formatting as desired.
% Do not put math or special symbols in the title.
\title{End-to-end translation of human neural activity to speech with a dual-dual generative adversarial network}
%
% author names and IEEE memberships
% note positions of commas and nonbreaking spaces ( ~ ) LaTeX will not break
% a structure at a ~ so this keeps an author's name from being broken across
% two lines.
% use \thanks{} to gain access to the first footnote area
% a separate \thanks must be used for each paragraph as LaTeX2e's \thanks
% was not built to handle multiple paragraphs
%

\author{Yina~Guo,~\IEEEmembership{Member,~IEEE,}
        Xiaofei~Zhang,~
        Zhenying~Gong,~
        Anhong~Wang
        and~Wenwu~Wang,~\IEEEmembership{Senior~Member,~IEEE}% <-this % stops a space
\thanks{This work was supported by National Natural Science Foundation of China under Grant 61301250, China Scholarship Council under Grant [2020]1417, Key Research and Development Project of Shanxi Province under Grant 201803D421035, Natural Science Foundation for Young Scientists of Shanxi Province under Grant 201901D211313, Shanxi Scholarship Council of China under Grant HGKY2019080 and
2020-127.}
\thanks{Yina Guo, Xiaofei Zhang, Zhenying Gong and Anhong Wang are with the School of Electronic Information Engineering, Taiyuan University of Science and Technology, Taiyuan 030024, China. Yina Guo and Xiaofei Zhang contributed equally to this work (Corresponding to: zulibest@tyust.edu.cn).}
\thanks{Wenwu Wang is with the Centre for Vision, Speech and Signal Processing, University of Surrey, Guildford, Surrey GU2 7XH, U.K. (Corresponding to: w.wang@surrey.ac.uk).}
\thanks{This work has been submitted to the IEEE for possible publication. Copyright may be transferred without notice, after which this version may no longer be accessible.}
% <-this % stops a space
%\thanks{Manuscript received January 19, 2017; revised March 26, 2017.}
}
\maketitle

% As a general rule, do not put math, special symbols or citations
% in the abstract or keywords.
\begin{abstract}
In a recent study of auditory evoked potential (AEP) based brain-computer interface (BCI), it was shown that, with an encoder-decoder framework, it is possible to translate human neural activity to speech (T-CAS). However, current encoder-decoder-based methods achieve T-CAS often with a two-step method where the information is passed between the encoder and decoder with a shared dimension reduction vector, which may result in a loss of information. A potential approach to this problem is to design an end-to-end method by using a dual generative adversarial network (DualGAN) without dimension reduction of passing information, but it cannot realize one-to-one signal-to-signal translation (see Fig. \ref{fig:1} (a) and (b)). In this paper, we propose an end-to-end model to translate human neural activity to speech directly, create a new electroencephalogram (EEG) datasets for participants with good attention by design a device to detect participants' attention, and introduce a dual-dual generative adversarial network (Dual-DualGAN) (see Fig. \ref{fig:1} (c) and (d)) to address an end-to-end translation of human neural activity to speech (ET-CAS) problem by group labelling EEG signals and speech signals, inserting a transition domain to realize cross-domain mapping. In the transition domain, the transition signals are cascaded by the corresponding EEG and speech signals in a certain proportion, which can build bridges for EEG and speech signals without corresponding features, and realize one-to-one cross-domain EEG-to-speech translation. The proposed method can translate word-length and sentence-length sequences of neural activity to speech. Experimental evaluation has been conducted to show that the proposed method significantly outperforms state-of-the-art methods on both words and sentences of auditory stimulus.
\end{abstract}

% Note that keywords are not normally used for peerreview papers.

% For peer review papers, you can put extra information on the cover
% page as needed:
% \ifCLASSOPTIONpeerreview
% \begin{center} \bfseries EDICS Category: 3-BBND \end{center}
% \fi
%
% For peerreview papers, this IEEEtran command inserts a page break and
% creates the second title. It will be ignored for other modes.
\IEEEpeerreviewmaketitle

\section{Introduction}
% The very first letter is a 2 line initial drop letter followed
% by the rest of the first word in caps.
%
% form to use if the first word consists of a single letter:
% \IEEEPARstart{A}{demo} file is ....
%
% form to use if you need the single drop letter followed by
% normal text (unknown if ever used by the IEEE):
% \IEEEPARstart{A}{}demo file is ....
%
% Some journals put the first two words in caps:
% \IEEEPARstart{T}{his demo} file is ....
%
% Here we have the typical use of a "T" for an initial drop letter
% and "HIS" in caps to complete the first word.
\IEEEPARstart{T}{he} World Health Organization (WHO) estimated in 2021 that neurological disorders could affect as many as 25$\%$ patients worldwide, and result in symptoms include confusion, altered levels of consciousness, and loss of communication. The visual evoked potential (VEP) based brain-computer interface (BCI) may enhance the quality of life of a patient, e.g. by using eyes to control a cursor to select letters one-by-one to spell out words
\cite{ryan2010predictive},
\cite{faller2012bcis},
\cite{manor2016multimodal}. However, the spelling rates of users are far below the average rate of 150 words per min in natural speech
\cite{jin2017improved},
\cite{norton2017elicitation}, since spelling is a sequential concatenation of discrete letters
\cite{guo2019investigation},
\cite{chailloux2020single},
\cite{bassi2021transfer}.
Different from spelling, speech is a highly efficient form of communication produced
from a fluid stream of overlapping and multi-articulator vocal tract movements
\cite{nijboer2008auditory},
\cite{klobassa2009toward}
\cite{kubler2009brain}. The auditory evoked potential (AEP) based BCI is a promising alternative to overcome the limitations of current spelling-based methods in achieving natural communication rates
\cite{brumberg2009artificial},
\cite{hohne2011novel},
\cite{hohne2012natural},
\cite{bocquelet2016real},
\cite{akbari2019towards},
\cite{anumanchipalli2019speech}.

\begin{figure*}[!t]
%\begin{figure}[!htb]
%\usepackage{float}
%\begin{figure*}[H]
\centering
\includegraphics[width=3.5in]{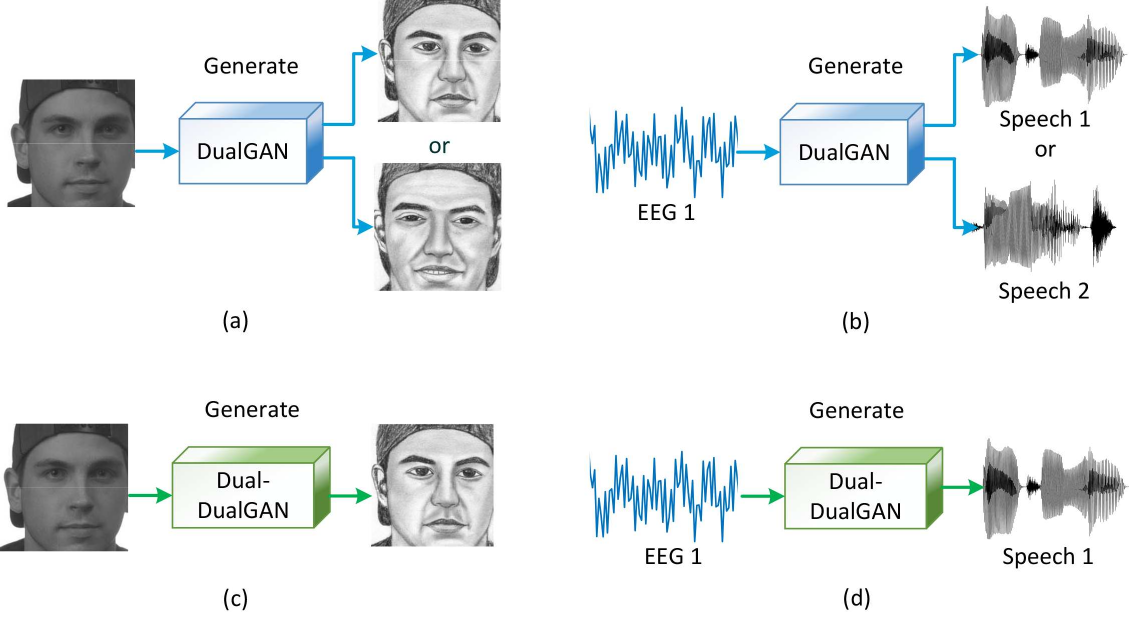}
\caption{Translation results of a DualGAN and our end-to-end model (Dual-DualGAN) (a) Image-to-image translation by a DualGAN (b) EEG-to-speech translation by a DualGAN (c) Image-to-image translation by a Dual-DualGAN (d) EEG-to-speech translation by a Dual-DualGAN.}
\label{fig:1}
\end{figure*}

The AEP based BCI using spelling-based methods, however, are yet to reach natural communication rates. To address this problem, studies have been conducted to exploit the conceptual similarity between the task of decoding speech from human neural activity and the task of machine translation according to the sensitivity of organs (such as ears, eyes). The AEP based BCI methods can be mainly classified into two categories: translation of human neural activity to text (T-CAT)
\cite{herff2015brain},
\cite{makin2020machine},
\cite{willett2020high} and translation of human neural activity to speech (T-CAS)
\cite{kim2011classification},
\cite{de2013lateralization},
\cite{hohne2014towards},
\cite{joos2014sensation},
\cite{yin2016auditory},
\cite{bocquelet2016real},
\cite{halder2016training},
\cite{heo2017music},
\cite{kaongoen2017novel},
\cite{hubner2018eyes},
\cite{huang2018usage}. The T-CAT method is mainly used by deaf-mutes. However, this method can be limited in several scenarios. For example, when two words share the same pronunciation, the translation to the desired word can be ambiguous. In addition, the spelling rates achieved by T-CAT can only be close to able-bodied typing rates. In contrast, the T-CAS method can be used by more ordinary users, as it is a more intuitive approach for communication, as in natural speech \cite{akbari2019towards},
\cite{anumanchipalli2019speech},
\cite{das2020stimulus},
\cite{krishna2020speech},
\cite{velasco2021speech}. As a result, the T-CAS approach has received increasing interest recently, and is the focus of this paper.

In existing AEP based BCIs, the T-CAS is often achieved with a two-step method, in which the first step is to decode speech from neural activity to text or acoustic feature, typically for dimensionality reduction, followed by a second step on encoding text or acoustic feature to synthesized speech. In the two-step method, a shared feature is a dimension reduction vector to bridge the decoding and encoding side, but dimensionality reduction inevitably leads to a loss of information. In encoding side, for a loss of some effective information, it may result in some possible reconstructed signals that are not necessarily accurate by using different encoders.
\cite{akbari2019towards},
\cite{anumanchipalli2019speech},
\cite{das2020stimulus},
\cite{velasco2021speech},
\cite{anumanchipalli2019speech}. To our knowledge, there is no existing study for end-to-end decoding human neural activity to speech by using a non-invasive electroencephalogram (EEG) neural recording without a loss of information for dimensionality reduction.

The aim of this paper is to develop a method for translating human neural activity to speech directly. Motivated by recent advances in dual generative adversarial network (DualGAN)
\cite{yi2017dualgan}, a potential solution to this problem is to design an end-to-end method by using a DualGAN without using dimensionality reduction in the pipeline, which, however, may be limited by the following challenges. For example, the DualGAN is an unsupervised dual learning framework originally designed for cross-domain image-to-image translation, but it cannot achieve a one-to-one translation for signal pairs without local corresponding features in signal pairs. As shown in Fig. \ref{fig:1} (a), a male photo may be translated to the corresponding male sketch or other similar male sketches by the DualGAN. This is because the image pairs have some correspondence in different patterns, for example hat or hair. In Fig. \ref{fig:1} (b), an EEG signal may be translated to different speech signals randomly by a DualGAN for the EEG and speech signals without local corresponding features (waveform and amplitude). In order to address these technical challenges, an end-to-end translation of human neural activity to speech (ET-CAS) problem is considered and our contributions are three-fold:
\begin{enumerate}
	\item \textbf{Model.} An end-to-end model is proposed to translate human neural activity to speech directly.
	\item \textbf{Datasets.} A new EEG dataset is created for this study, where a device (see Fig. \ref{fig:7}) is designed to detect the attention of participants in order to improve the quality of the EEG data in data collection.
	\item \textbf{Network.} A dual-dual generative adversarial network (Dual-DualGAN) is proposed to address the ET-CAS problem, where two DualGANs are built and trained simultaneously by incorporating a transition domain to bridge the two DualGANs. The transition signals used in the transition domain are cascaded by the corresponding EEG and speech signals in a certain proportion to construct shared labels for EEG and speech signals without aligning their corresponding features.
\end{enumerate}

The remainder of the paper is organized as follows. Section II describes the related work. Section III introduces the background for GAN and DualGAN. Section IV formulates an end-to-end model for the ET-CAS problem. Section V presents our proposed network for the problem of ET-CAS. Section VI describes data collection and pre-processing. Section VII discuses the experimental set up. Section VIII shows numerical results. Section ¢ù concludes the paper and draws potential future research directions.

\begin{figure*}[!t]
%\begin{figure}[!htb]
%\usepackage{float}
%\begin{figure*}[H]
\centering
\includegraphics[width=5in]{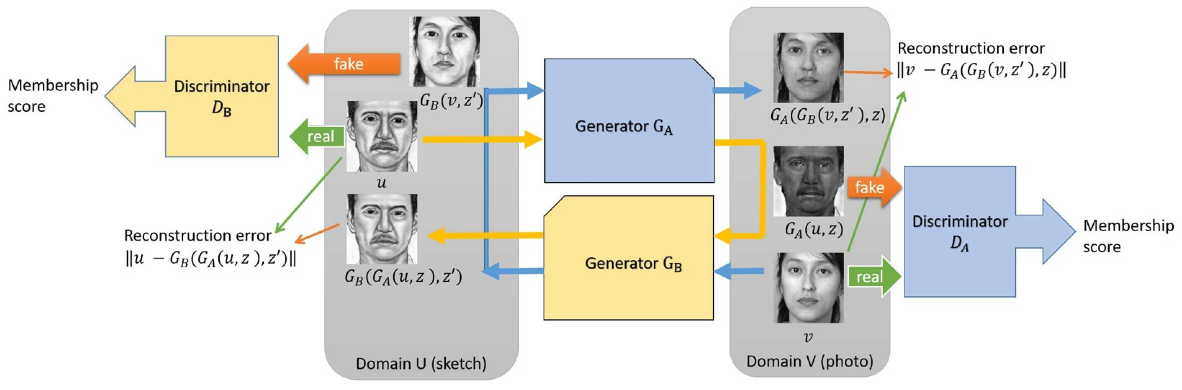}
\caption{Architecture and data flow chart of DualGAN for image-to-image translation.}
\label{fig:2}
\end{figure*}
\section{Related Work}
In the field of the AEP based BCIs, there is an increasing interest in the problem of decoding speech from human neural activity recently \cite{akbari2019towards},
\cite{anumanchipalli2019speech},
\cite{das2020stimulus},
\cite{krishna2020speech},
\cite{velasco2021speech}. According to the sensitivity of organs (e.g. ears, eyes), the AEP based BCI systems can be mainly classified into two categories, namely, T-CAT and T-CAS. The T-CAT systems are more suitable for deaf and mute people, and the spelling rates offered by these systems are close to typing rates. However, they are prone to errors when translating two words with the same pronunciation. In contrast, the T-CAS method does not have such limitations, and is an intuitive approach for users to achieve high communication rates as in natural speech.

\textbf{The AEP based BCIs for T-CAT.} Herff et al.
\cite{herff2015brain} showed for the first time that spoken speech could be decoded into the expressed words from intracranial electrocorticographic (ECoG) recordings, and proposed a Brain-To-Text system to transform brain activity while speaking into the corresponding textual representation. This system achieved word error rates as low as 25 $\%$ and phone error rates below 50 $\%$. Makin et al.
\cite{makin2020machine} trained a recurrent neural network to encode each sentence-long sequence of neural activity into an abstract representation, and then to decode this representation, word by word, into an English sentence at natural-speech rates with high accuracy. This method achieved an average word error rate across a held-out repeat set as low as 3 $\%$. Willett et al.
\cite{willett2020high} proposed a BCI which can spell 90 characters per minute at $>$ 99 $\%$ accuracy with general-purpose auto-correction, and significantly close the gap between BCI-enabled typing and able-bodied typing rates. These methods have focused on the translation of human neural activity to text and achieved typing rates that are close to normal typing rates, which, however, remain lower than the average rate of 150 words per min in natural speech.

\textbf{The AEP based BCIs for T-CAS.} Brumberg et al.
\cite{brumberg2009artificial} developed a brain-computer interface for the control of an artificial speech synthesizer by an individual with near complete paralysis, where vowel formant frequencies are predicted based on neural activity recorded from an intra-neural micro-electrode implanted in the left hemisphere speech motor cortex.
Bocquelet et al.
\cite{bocquelet2016real} presented an articulatory-based speech synthesizer, which converts movements of the main speech articulators (e.g. tongue, jaw, velum, and lips) into intelligible speech by using a deep neural network (DNN), and can be controlled in real-time that is useful for BCI applications. Akbari et al.
\cite{akbari2019towards} investigated the dependence of reconstruction accuracy on linear and nonlinear (deep neural network) regression methods and the acoustic representation. Anumanchipalli et al.
\cite{anumanchipalli2019speech} designed a neural decoder that explicitly leverages kinematic and sound representations encoded in human neural activity to synthesize audible speech. These methods demonstrated the possibility to translate human neural activity to speech with encoder-decoder frameworks. However, there are two major open challenges for T-CAS. First, the collection of intracranial ECoG recordings is intrusive and inconvenient. Second, the encoder-decoder based methods need multi-steps to achieve T-CAS. Krishna et al.
\cite{krishna2020speech} demonstrated synthesizing speech from the non-invasive electroencephalogram (EEG) neural recordings for the first time and proposed a recurrent neural network (RNN) regression model to predict mel-frequency cepstral coefficients (MFCC) from EEG features. This method shows that it is possible to decode the human neural activity by non-invasive EEG neural recordings, but doesn't consider the speech reconstruction from EEG features.

The focus of this paper is to address the problem of decoding speech from human neural activity directly with the non-invasive EEG signals, and to propose an end-to-end translation method Dual-DualGAN.

\section{Background}
A generative adversarial network (GAN) is a class of machine learning frameworks designed by Goodfellow et al.
\cite{2014Generative}. Given a training set, a GAN learns to generate new data with the same statistics as the training set. For example, a GAN trained on photos can generate new photos that look at least superficially authentic to human observers, having many realistic characteristics. However, the original GAN algorithm is not directly applicable for the translation task for training network without paired signals, and the trained photos and the generated photos without one-to-one correspondence.

Motivated by GAN and dual learning, an unsupervised learning framework DualGAN has been proposed by Yi et al.
\cite{yi2017dualgan} for image translation, which was trained with two sets of unlabeled images from two domains (e.g. sketch and photo).

As illustrated in Fig. \ref{fig:2}, given two sets of unlabeled and unpaired images sampled from domains $U$ and $V$, respectively, the primary task of DualGAN is to learn a generator $G_{A}$: $U\rightarrow V$ that maps an image $\textrm{\textbf{u}} \in U$ to an image $\textrm{\textbf{v}} \in V$, while the dual task is to train an inverse generator $G_{B}$: $V\rightarrow U$. This is achieved by using two GANs (i.e. the primal GAN and the dual GAN). The primal GAN learns the generator $G_{A}$ and a discriminator $D_{A}$ that discriminates between fake outputs of $G_{A}$ and real members of domain $V$. Analogously, the dual GAN learns the generator $G_{B}$ and a discriminator $D_{B}$.

The image $\textrm{\textbf{u}} \in U$ is translated to domain $V$ using $G_{A}$. How well the translation $G_{A}(\textrm{\textbf{u}},\textrm{\textbf{z}})$ fits in $V$ is evaluated by $D_{A}$, where $\textrm{\textbf{z}}$ is random noise. $G_{A}(\textrm{\textbf{u}},\textrm{\textbf{z}})$ is then translated back to domain $U$ using $G_{A}$, which outputs $G_{B}(G_{A}(\textrm{\textbf{u}},\textrm{\textbf{z}}),\textrm{\textbf{z}}')$ as the reconstructed version of $\textrm{\textbf{u}}$, where $\textrm{\textbf{z}}'$ is also random noise. Similarly, $\textrm{\textbf{v}} \in V$ is translated to $U$ as $G_{B}(\textrm{\textbf{v}},\textrm{\textbf{z}}')$ and then reconstructed as $G_{A}(G_{B}(\textrm{\textbf{v}},\textrm{\textbf{z}}'),\textrm{\textbf{z}})$.
The discriminator $D_{A}$ is trained with $\textrm{\textbf{v}}$ as positive examples
and $G_{A}(\textrm{\textbf{u}},\textrm{\textbf{z}})$ as negative examples, whereas $D_{B}$ takes $\textrm{\textbf{u}}$ as positive and $G_{B}(\textrm{\textbf{v}},\textrm{\textbf{z}}')$ as negative. Generators $G_{A}$ and $G_{B}$ are optimized to emulate ¡°fake¡± outputs to blind the corresponding discriminators $D_{A}$ and $D_{B}$, as well as to minimize the two reconstruction losses $\|\textrm{\textbf{u}}-G_{B}(G_{A}(\textrm{\textbf{u}},\textrm{\textbf{z}}),\textrm{\textbf{z}}')\|$ and $\|\textrm{\textbf{v}}-G_{A}(G_{B}(\textrm{\textbf{v}},\textrm{\textbf{z}}'),\textrm{\textbf{z}})\|$.

The same loss function is used for both generators $G_{A}$ and $G_{B}$ as they share the same task, which is defined as
\begin{multline}
\mathcal{L}(U,V)^{G}=\lambda_{U}\|\textrm{\textbf{u}}-G_{B}(G_{A}(\textrm{\textbf{u}},\textrm{\textbf{z}}),\textrm{\textbf{z}}')\|\\
+\lambda_{V}\|\textrm{\textbf{v}}-G_{A}(G_{B}(\textrm{\textbf{v}},\textrm{\textbf{z}}'),\textrm{\textbf{z}})\| \quad\\
-D_{A}(G_{A}(\textrm{\textbf{u}},\textrm{\textbf{z}}))-D_{B}(G_{B}(\textrm{\textbf{v}},\textrm{\textbf{z}}')), \qquad\quad\
\label{eqs:1}
%\notag
\end{multline}
where $\lambda_{U}$ and $\lambda_{V}$ are two constant parameters, which are typically
set to the values within $[100,1000]$
\cite{yi2017dualgan}.

\begin{figure*}[!t]
%\begin{figure}[!htb]
%\usepackage{float}
%\begin{figure*}[H]
\centering
\includegraphics[width=5.5in]{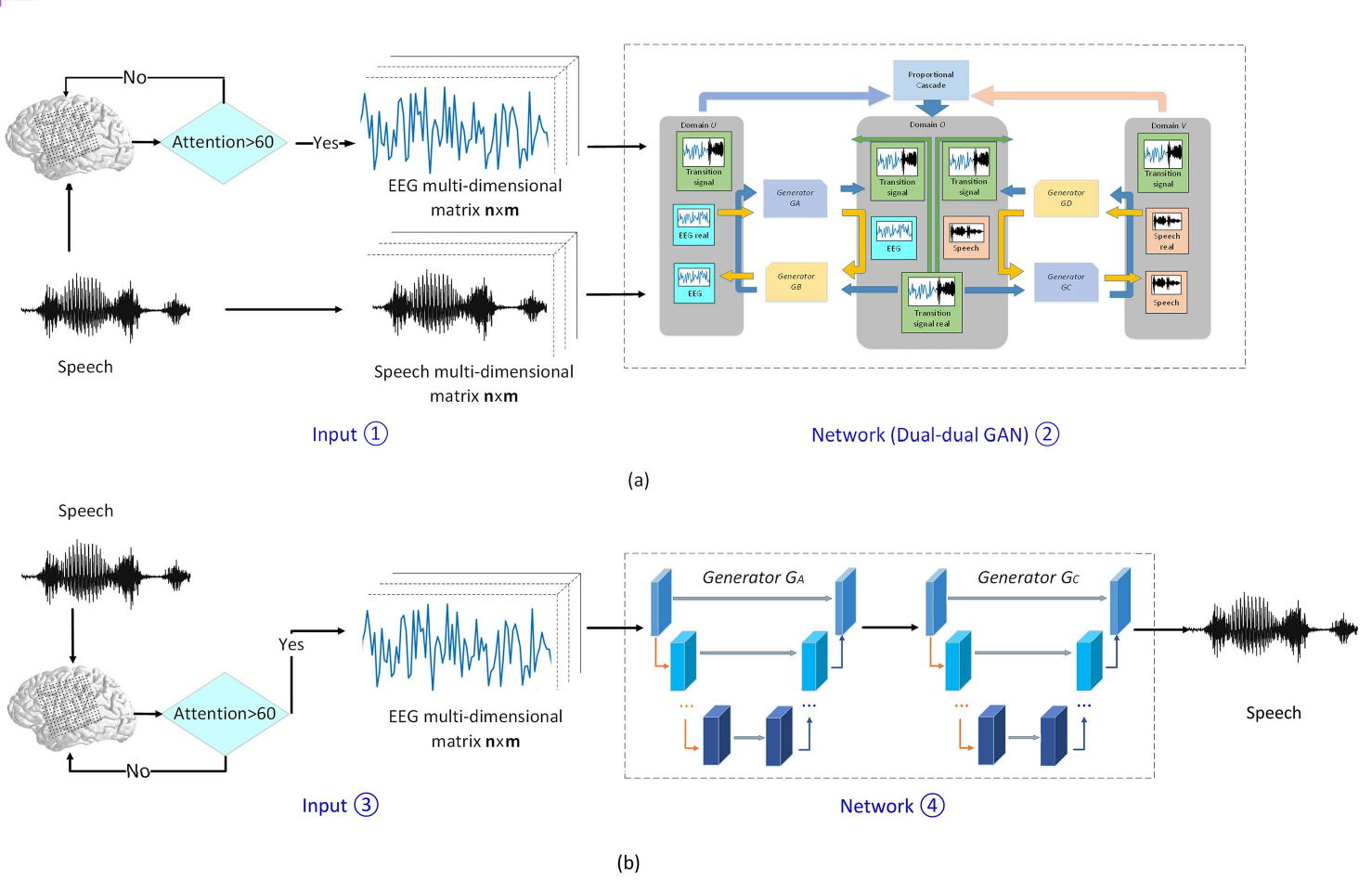}
\caption{Architecture of our end-to-end model (a) Training, including input \textcircled{1}, network (Dual-DualGAN) \textcircled{2} (b) Testing, including input \textcircled{3}, network \textcircled{4}.}
\label{fig:3}
\end{figure*}

The loss functions of $D_{A}$ and $D_{B}$ advocated by Wasserstein GAN (WGAN)
\cite{yi2017dualgan} can be described by
\begin{equation}
\mathcal{L}_{A}^{D}=D_{A}(G_{A}(\textrm{\textbf{u}},\textrm{\textbf{z}}))-D_{A}(\textrm{\textbf{v}}),
\label{eqs:2}
\end{equation}
\begin{equation}
\mathcal{L}_{B}^{D}=D_{B}(G_{B}(\textrm{\textbf{v}},\textrm{\textbf{z}}'))-D_{B}(\textrm{\textbf{u}}),
\label{eqs:3}
\end{equation}
where $D_{A}(\cdot)=\frac{p_{data}(\cdot)}{p_{data}(\cdot)+p_{gA}(\cdot)}$, $D_{B}(\cdot)=\frac{p_{data}(\cdot)}{p_{data}(\cdot)+p_{gB}(\cdot)}$ $p_{data}(\cdot)$ is the distribution of the training data, $p_{gA}(\cdot)$ and $p_{gB}(\cdot)$ are the distributions of the fake outputs from $G_{A}(\cdot)$ and $G_{B}(\cdot)$, respectively.
After several steps of training, if $p_{g}(\cdot)=p_{data}(\cdot)$, the discriminator is unable to differentiate between $p_{g}(\cdot)$ and $p_{data}(\cdot)$, and $D_{A}(\cdot)=D_{B}(\cdot)=\frac{1}{2}$.

By training a DualGAN, a signal can be translated to another similar signal with some correspondence in different patterns. For example, a male photo may be translated to the
corresponding male sketch or other similar male sketches by a DualGAN in Fig. \ref{fig:1} (a). However, the male photo and the translated male sketch are not necessarily the trained image pairs. Analogously, an EEG signal may be translated to some different speech signals randomly by a DualGAN in Fig. \ref{fig:1} (b). The correct rates of the one-to-one translation in EEG-to-speech are even lower than image pairs for the EEG and speech signals without local corresponding features.

For ET-CAS problem, we need to realize one-to-one translation of an EEG to a speech signal. To address this problem, in this paper, we build an end-to-end model for decoding speech from human neural activity to synthesized speech directly. Based on the end-to-end model, we propose a Dual-DualGAN by group labelling EEG signals and speech signals, and inserting a transition domain into a DualGAN to train two DualGAN simultaneously. The transition signals of the transition domain can be considered as the shared labels for the EEG and speech signals which are constructed by cascading the corresponding EEG and speech signals in a certain proportion.

\section{End-to-End Model}
Consider a translation task that is induced by decoding speech from human neural activity to synthesized speech directly, we present an end-to-end model for this task in this section, as illustrated in Fig. \ref{fig:3}.

\textbf{Training.} Fig. \ref{fig:3} (a) shows the training process of our end-to-end model. Two inputs are used in this model, including the non-invasive EEG signal recorded as the participants listen to speech audio, and the corresponding speech signal. We use these two inputs to train the proposed Dual-DualGAN, and then obtain the trained Dual-DualGAN.

\textbf{Testing.} Fig. \ref{fig:3} (b) demonstrates the testing process of the end-to-end model. The input of the model is the non-invasive EEG signal recorded as the participants listen to speech audio. The parameters derived from the trained Dual-DualGAN is used to decode EEG to the corresponding synthesized speech signal.

\section{Dual-DualGAN}
\begin{figure*}[!t]
%\begin{figure}[!htb]
%\usepackage{float}
%\begin{figure*}[H]
\centering
\includegraphics[width=6.5in]{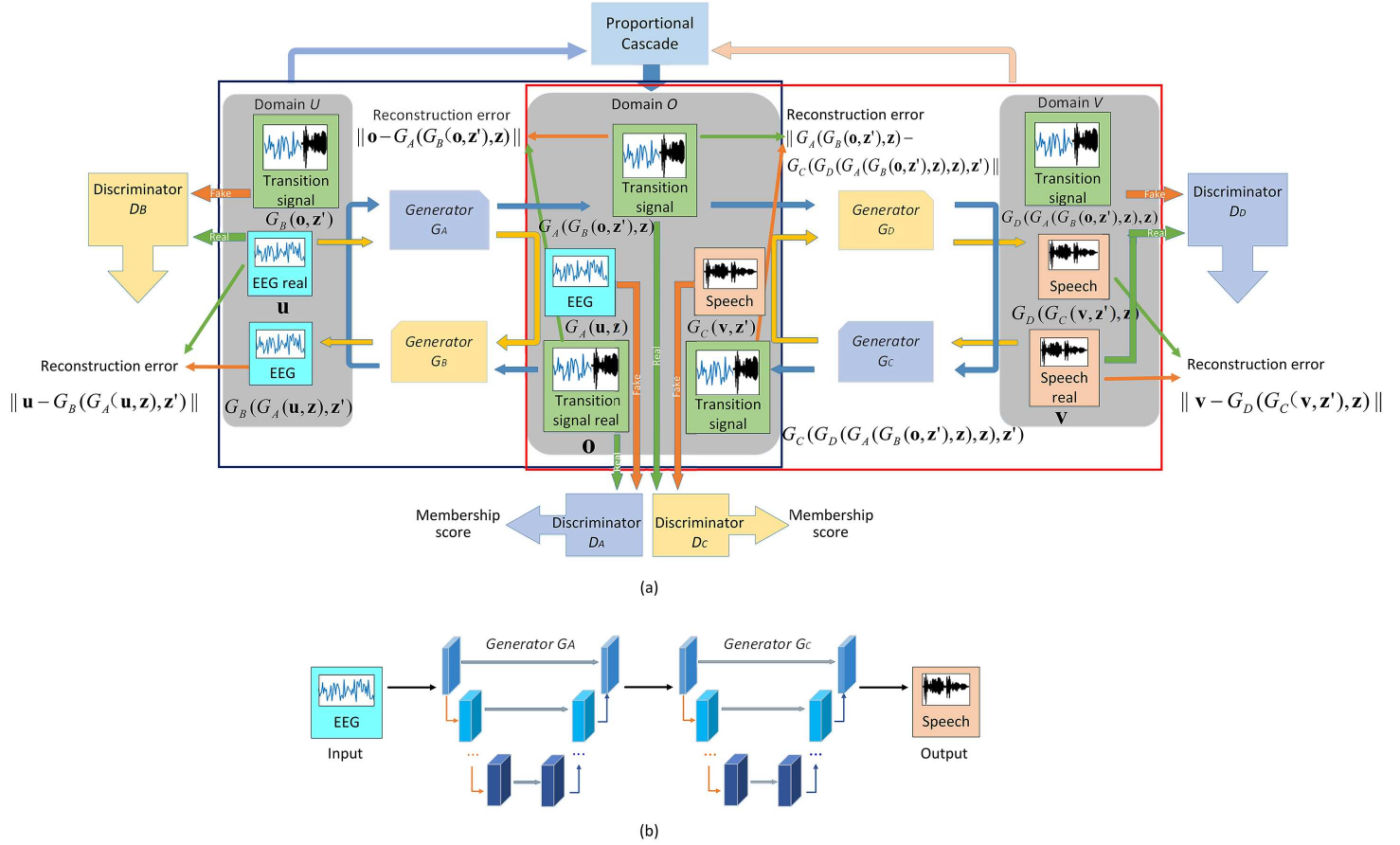}
\caption{Architecture of our network (a) Dual-DualGAN for training, including two DualGANs trained simultaneously. DualGAN 1 learns the mapping between the EEG signals $\textrm{\textbf{u}}\in U$ and the transition signals $\textrm{\textbf{o}}\in O$, while DualGAN 2 learns the mapping between the speech signals $\textrm{\textbf{v}}\in U$ and the generated transition signals $\textrm{\textbf{o}}\in O$. Thus we can find the mapping between the EEG signals $\textrm{\textbf{u}}\in U$ and the speech signals $\textrm{\textbf{v}}\in V$ to address the ET-CAS problem. (b) Network for testing, where the trained Dual-DualGAN is used to realize one-to-one EEG-to-speech translation.}
\label{fig:4}
\end{figure*}

The fundamental question in ET-CAS is to decode speech from human neural activity (EEG signals) directly. A potential approach to this problem is to use a DualGAN which is an unsupervised dual learning framework that does not need dimensionality reduction in cross-modal translation. However, it cannot realize one-to-one translation for signal pairs without local corresponding features in the signal pairs. The EEG and speech signals are different types of signals without local corresponding features. Thus we cannot achieve the purpose of the one-to-one translation of an EEG to a speech signal by training a DualGAN. To address this problem, we present a Dual-DualGAN, as shown in Fig. \ref{fig:4}.

\begin{figure*}[!t]
%\begin{figure}[!htb]
%\usepackage{float}
%\begin{figure*}[H]
\centering
\includegraphics[width=5in]{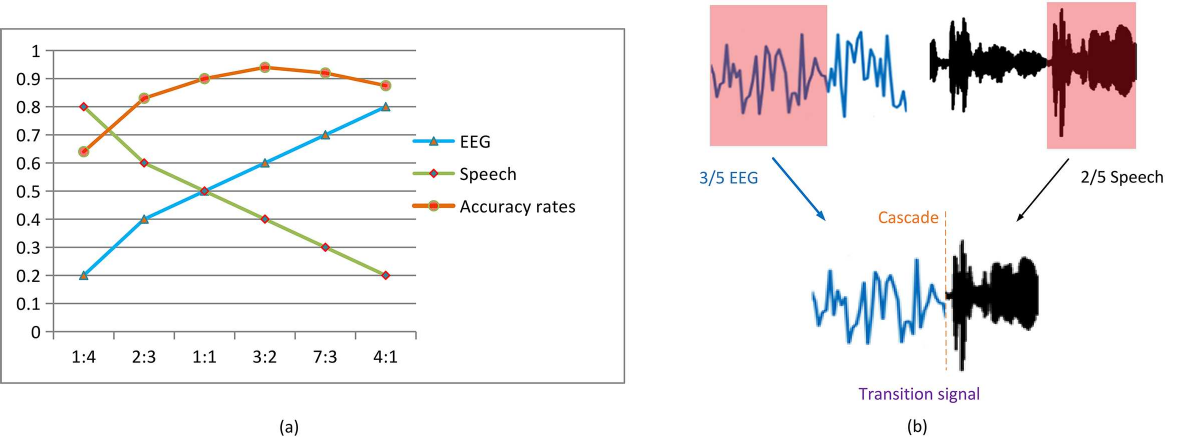}
\caption{Proportional cascade of the transition signals (a) Accuracy rates with different cascading proportion (b) A sample transition signal formed by cascading the EEG and speech signal in a proportion of 3 to 2.}
\label{fig:5}
\end{figure*}

In our Dual-DualGAN, a transition domain $O$ is introduced into a DualGAN to build two DualGANs (DualGAN 1 and DualGAN 2) which are trained simultaneously. DualGAN 1 involves domain $U$ and domain $O$ whereas DualGAN 2 involves domain $O$ and domain $V$. As illustrated in Fig. \ref{fig:5} (b), a set of transition signals $\textrm{\textbf{o}}$ sampled from $O$ are cascaded by the corresponding EEG and speech signals in a certain proportion. Two sets of EEG signals $\textrm{\textbf{u}}$ and speech signals $\textrm{\textbf{v}}$ are sampled from domains $U$ and $V$, respectively. The primary task of our Dual-DualGAN is to translate $\textrm{\textbf{u}}\in U$ into $\textrm{\textbf{v}}\in V$. DualGAN 1 aims to learn a mapping between the EEG signals $\textrm{\textbf{u}}\in U$ and transition signals $\textrm{\textbf{o}}\in O$ with a mini-cycle, while DualGAN 2 learns a mapping between the speech signals $\textrm{\textbf{v}}\in V$ and transition signals with another mini-cycle. Different from the above running mode, the transition signals $\textrm{\textbf{o}}\in O$ learn from $\textrm{\textbf{u}}\in U$ in DualGAN 1, then the generated transition signals learn from $\textrm{\textbf{v}}\in V$ in DualGAN 2, which form a large cycle. By training the Dual-DualGAN, the transition signals $\textrm{\textbf{o}}\in O$ can be considered as shared labels for EEG and speech signals without corresponding features, details as follows.

\textbf{Training.} In Fig. \ref{fig:4} (a), a generator $G_{A}$: $U\rightarrow O$ in the DualGAN 1 is learned by mapping a real EEG signal $\textrm{\textbf{u}}$ to an EEG signal $G_{A}(\textrm{\textbf{u}},\textrm{\textbf{z}})$, while the dual task is to train an inverse generator $G_{B}$: $O\rightarrow U$ that maps a real transition signal $\textrm{\textbf{o}}$ to a transition signal $G_{B}(\textrm{\textbf{o}},\textrm{\textbf{z}}')$. Analogously, a generator $G_{C}$: $V\rightarrow O$ in the DualGAN 2 is learned by mapping a real speech signal $\textrm{\textbf{v}}$ to a speech signal $G_{C}(\textrm{\textbf{v}},\textrm{\textbf{z}}')$, while the dual task is to train an inverse generator $G_{D}$: $O\rightarrow V$ that maps a generated transition signal of DualGAN 1 $G_{A}(G_{B}(\textrm{\textbf{o}},\textrm{\textbf{z}}'),\textrm{\textbf{z}})$ to a transition signal $G_{D}(G_{A}(G_{B}(\textrm{\textbf{o}},\textrm{\textbf{z}}'),\textrm{\textbf{z}}),\textrm{\textbf{z}})$, where $\textrm{\textbf{z}}$ and $\textrm{\textbf{z}}'$ are random noises.

For EEG signals, a real EEG signal $\textrm{\textbf{u}}$ is mapped to domain
$O$ using $G_{A}$, which generates an EEG signal $G_{A}(\textrm{\textbf{u}},\textrm{\textbf{z}})$. Then $G_{A}(\textrm{\textbf{u}},\textrm{\textbf{z}})$ is translated back to domain
$U$ using $G_{B}$, which outputs $G_{B}(G_{A}(\textrm{\textbf{u}},\textrm{\textbf{z}}),\textrm{\textbf{z}}')$ as the reconstructed version of $\textrm{\textbf{u}}$.

For speech signals, a real speech signal $\textrm{\textbf{v}}$ is mapped to domain
$O$ using $G_{C}$, which generates a speech signal $G_{C}(\textrm{\textbf{v}},\textrm{\textbf{z}}')$. $G_{C}(\textrm{\textbf{v}},\textrm{\textbf{z}}')$ is then translated back to domain
$V$ using $G_{D}$, which outputs $G_{D}(G_{C}(\textrm{\textbf{v}},\textrm{\textbf{z}}'),\textrm{\textbf{z}})$ as the reconstructed version of $\textrm{\textbf{v}}$.

For transition signals, it needs four steps to form a large cycle. Firstly, a real transition signal $\textrm{\textbf{o}}$ is mapped to domain
$U$ using $G_{B}$, which generates a transition signal $G_{B}(\textrm{\textbf{o}},\textrm{\textbf{z}}')$. Secondly, $G_{B}(\textrm{\textbf{o}},\textrm{\textbf{z}}')$ is translated back to domain
$O$ using $G_{A}$, which outputs $G_{A}(G_{B}(\textrm{\textbf{o}},\textrm{\textbf{z}}'),\textrm{\textbf{z}})$. Thirdly, $G_{A}(G_{B}(\textrm{\textbf{o}},\textrm{\textbf{z}}'),\textrm{\textbf{z}})$ is translated to domain $V$ using $G_{D}$, which generates $G_{D}(G_{A}(G_{B}(\textrm{\textbf{o}},\textrm{\textbf{z}}'),\textrm{\textbf{z}}),\textrm{\textbf{z}})$. Fourthly, $G_{D}(G_{A}(G_{B}(\textrm{\textbf{o}},\textrm{\textbf{z}}'),\textrm{\textbf{z}}),\textrm{\textbf{z}})$ is translated back to domain $O$ using $G_{C}$, which outputs $G_{C}(G_{D}(G_{A}(G_{B}(\textrm{\textbf{o}},\textrm{\textbf{z}}'),\textrm{\textbf{z}}),\textrm{\textbf{z}}),\textrm{\textbf{z}}')$
as the reconstructed version of $\textrm{\textbf{o}}$.

The discriminator $D_{A}$ in DualGAN 1 is learned by discriminating between the real transition signal $\textrm{\textbf{o}}$ of domain $O$ and the fake outputs of $G_{A}$, while the discriminator $D_{B}$ is learned by discriminating between the real EEG signal $\textrm{\textbf{u}}$ of domain $U$ and the fake outputs of $G_{B}$. Similarly, the discriminator $D_{D}$ in DualGAN 2 is learned by discriminating between the real speech signal $\textrm{\textbf{v}}$ of domain $V$ and the fake outputs of $G_{D}$, while the discriminator $D_{C}$ is learned by discriminating between the generated transition signal of DualGAN 1 $G_{A}(G_{B}(\textrm{\textbf{o}},\textrm{\textbf{z}}'),\textrm{\textbf{z}})$ of domain $O$ and the fake outputs of $G_{C}$.

The generators $G_{A}$, $G_{B}$, $G_{C}$ and $G_{D}$
are optimized to emulate the fake outputs to blind the corresponding discriminators $D_{A}$, $D_{B}$, $D_{C}$ and $D_{D}$ as well as to minimize the following reconstruction losses $\|\textrm{\textbf{u}}-G_{B}(G_{A}(\textrm{\textbf{u}},\textrm{\textbf{z}}),\textrm{\textbf{z}}')\|$, $\|\textrm{\textbf{o}}-G_{A}(G_{B}(\textrm{\textbf{o}},\textrm{\textbf{z}}'),\textrm{\textbf{z}})\|$,
$\|\textrm{\textbf{v}}-G_{D}(G_{C}(\textrm{\textbf{v}},\textrm{\textbf{z}}'),\textrm{\textbf{z}})\|$, and $\|G_{A}(G_{B}(\textrm{\textbf{o}},\textrm{\textbf{z}}'),\textrm{\textbf{z}})-G_{C}(G_{D}(G_{A}(G_{B}(\textrm{\textbf{o}},\textrm{\textbf{z}}'),\textrm{\textbf{z}}),\textrm{\textbf{z}}),\textrm{\textbf{z}}')\|$.

The same loss function is used in DualGAN 1 for generators $G_{A}$ and $G_{B}$ as they share the same task
\cite{yi2017dualgan}, which is defined as
\begin{multline}
\mathcal{L}(U,O)^{G}=\lambda_{U}\|\textrm{\textbf{u}}-G_{B}(G_{A}(\textrm{\textbf{u}},\textrm{\textbf{z}}),\textrm{\textbf{z}}')\|\\
+\lambda_{O}\|\textrm{\textbf{o}}-G_{A}(G_{B}(\textrm{\textbf{o}},\textrm{\textbf{z}}'),\textrm{\textbf{z}})\|\\
\quad\ \
-D_{A}(G_{A}(\textrm{\textbf{u}},\textrm{\textbf{z}}))-D_{B}(G_{B}(\textrm{\textbf{o}},\textrm{\textbf{z}}')),\ \quad\
\label{eqs:4}
\end{multline}
where $\lambda_{U}$ and $\lambda_{O}$ are two constant parameters, which are typically
set to the values within $[100,1000]$
\cite{yi2017dualgan}.

Analogously, the loss function in DualGAN 2 for both generators $G_{C}$ and $G_{D}$ is defined as follows
\begin{multline}
\mathcal{L}(V,O)^{G}=\lambda_{V}\|\textrm{\textbf{v}}-G_{D}(G_{C}(\textrm{\textbf{v}},\textrm{\textbf{z}}'),\textrm{\textbf{z}})\|\\
+\lambda_{O}\|G_{A}(G_{B}(\textrm{\textbf{o}},\textrm{\textbf{z}}'),\textrm{\textbf{z}})-G_{C}(G_{D}(G_{A}(G_{B}(\textrm{\textbf{o}},\textrm{\textbf{z}}'),\textrm{\textbf{z}}),\textrm{\textbf{z}}),\textrm{\textbf{z}}')\|\\
-D_{D}(G_{D}(G_{A}(G_{B}(\textrm{\textbf{o}},\textrm{\textbf{z}}'),\textrm{\textbf{z}}),\textrm{\textbf{z}}))-D_{C}(G_{C}(\textrm{\textbf{v}},\textrm{\textbf{z}}')),\ \qquad\
\label{eqs:5}
\end{multline}
where $\lambda_{V}$ is a constant parameter which can be set typically to the values within $[100,1000]$.

The loss functions of $D_{A}$, $D_{B}$, $D_{C}$ and $D_{D}$ advocated by Wasserstein GAN (WGAN)
\cite{yi2017dualgan} can be described by
\begin{equation}
\mathcal{L}_{A}^{D}=D_{A}(G_{A}(\textrm{\textbf{u}},\textrm{\textbf{z}}))-D_{A}(\textrm{\textbf{o}}),
\label{eqs:6}
\end{equation}
\begin{equation}
\mathcal{L}_{B}^{D}=D_{B}(G_{B}(\textrm{\textbf{o}},\textrm{\textbf{z}}'))-D_{B}(\textrm{\textbf{u}}),
\label{eqs:7}
\end{equation}
\begin{equation}
\mathcal{L}_{C}^{D}=D_{C}(G_{C}(\textrm{\textbf{v}},\textrm{\textbf{z}}'))-D_{C}(G_{A}(G_{B}(\textrm{\textbf{o}},\textrm{\textbf{z}}'),\textrm{\textbf{z}})),
\label{eqs:8}
\end{equation}
\begin{equation}
\mathcal{L}_{D}^{D}=D_{D}(G_{D}(G_{A}(G_{B}(\textrm{\textbf{o}},\textrm{\textbf{z}}'),\textrm{\textbf{z}}),\textrm{\textbf{z}}))-D_{D}(\textrm{\textbf{v}}),
\label{eqs:9}
\end{equation}
where $D_{A}(\cdot)$, $D_{B}(\cdot)$, $D_{C}(\cdot)$ and $D_{D}(\cdot)$ are defined similarly as in \eqref{eqs:2} and \eqref{eqs:3}.

With adversarial training of the proposed Dual-DualGAN, we find the mapping between the EEG signals $\textrm{\textbf{u}}\in U$ and the transition signals $\textrm{\textbf{o}}\in O$, and the mapping between the speech signals $\textrm{\textbf{v}}\in V$ and the transition signals $\textrm{\textbf{o}}\in O$. Thus the
transition signals $\textrm{\textbf{o}}\in O$ can be considered as shared labels for the EEG signals $\textrm{\textbf{u}}\in U$ and the speech signals $\textrm{\textbf{v}}\in V$ without corresponding features, which facilitates the one-to-one translation from EEG to speech.

\textbf{Testing.} As shown in Fig. \ref{fig:4} (b), with adversarial training of the proposed Dual-DualGAN, we can obtain the trained parameters of the Dual-DualGAN. The EEG signals $\textrm{\textbf{u}}\in U$ as the inputs can be translated to the speech signals $\textrm{\textbf{v}}\in V$ by using the trained parameters of the Dual-DualGAN, which realize one-to-one EEG-to-speech translation.

\textbf{Network configuration.} Fig. \ref{fig:4} (a) includes three domains: the domain $U$ with the EEG signals $\textrm{\textbf{u}}$, the domain $V$ with the speech signals $\textrm{\textbf{v}}$, and the transition domain $O$ with the transition signals $\textrm{\textbf{o}}$. By inserting the transition domain $O$ with the transition signals $\textrm{\textbf{o}}$, the cross-domain EEG-to-speech translation can be realized. The transition signals $\textrm{\textbf{o}}$ are obtained by cascading the corresponding EEG and speech signals in a certain proportion. As illustrated in Fig. \ref{fig:5}, the EEG and speech signals are cascaded in different proportions from 1 : 4 to 4 : 1 with the step of $\frac{1}{5}$. For low values of the proportion 1 : 4 and 2 : 3, the accuracy rates are relatively low (at around 0.63 and 0.82, respectively). When the values of the proportion is higher than 1 : 1, the accuracy rate can rise above 0.88, in which the accuracy rate reaches the highest value 0.95 in a proportion of 3 to 2. Thus we choose the proportion of 3:2 to cascade the EEG and speech signals in this paper.

The proposed network is summarized in Algorithm \ref{alg:1}.
\begin{algorithm}[htbp]
\caption{Dual-DualGAN}
\label{alg:1}
\begin{algorithmic}%[1]
\Require
EEG signals $\textrm{\textbf{u}}\in U$, speech signals $\textrm{\textbf{v}}\in V$, transition signals $\textrm{\textbf{o}}\in O$, the number of critic iterations per generator iteration $N$, $\lambda_{U}$, $\lambda_{U}$, $\lambda_{O}$, an initial learning rate, and batch size $M$, which are depicted in Section V.
\Ensure
One-to-one EEG-to-speech translation of $\textrm{\textbf{u}}\in U$ to $\textrm{\textbf{v}}\in V$.
\State 1.
\textit{Loss function of generators $G_{A}$ and $G_{B}$ in DualGAN 1}.\\
$\mathcal{L}(U,O)^{G}=\lambda_{U}\|\textrm{\textbf{u}}-G_{B}(G_{A}(\textrm{\textbf{u}},\textrm{\textbf{z}}),\textrm{\textbf{z}}')\|$
\\
$+\lambda_{O}\|\textrm{\textbf{o}}-G_{A}(G_{B}(\textrm{\textbf{o}},\textrm{\textbf{z}}'),\textrm{\textbf{z}})\|$
\\
$-D_{A}(G_{A}(\textrm{\textbf{u}},\textrm{\textbf{z}}))-D_{B}(G_{B}(\textrm{\textbf{o}},\textrm{\textbf{z}}')),\ \quad$
\\
where $\lambda_{U}$ and $\lambda_{O}$ are defined as in \eqref{eqs:4}.
\State 2.
\textit{Loss function of generators $G_{C}$ and $G_{D}$ in DualGAN 2}.\\
$\mathcal{L}(V,O)^{G}=\lambda_{V}\|\textrm{\textbf{v}}-G_{D}(G_{C}(\textrm{\textbf{v}},\textrm{\textbf{z}}'),\textrm{\textbf{z}})\|$
\\
$+\lambda_{O}\|G_{A}(G_{B}(\textrm{\textbf{o}},\textrm{\textbf{z}}'),\textrm{\textbf{z}})$
\\
$-G_{C}(G_{D}(G_{A}(G_{B}(\textrm{\textbf{o}},\textrm{\textbf{z}}'),\textrm{\textbf{z}}),\textrm{\textbf{z}}),\textrm{\textbf{z}}')\|$
\\
$-D_{D}(G_{D}(G_{A}(G_{B}(\textrm{\textbf{o}},\textrm{\textbf{z}}'),\textrm{\textbf{z}}),\textrm{\textbf{z}}))-D_{C}(G_{C}(\textrm{\textbf{v}},\textrm{\textbf{z}}')),$
\\
where $\lambda_{V}$ is defined as in \eqref{eqs:5}.
\State 3. \textit{Loss function of discriminators $D_{A}$, $D_{B}$, $D_{C}$ and $D_{D}$}.\\
$\mathcal{L}_{A}^{D}=D_{A}(G_{A}(\textrm{\textbf{u}},\textrm{\textbf{z}}))-D_{A}(\textrm{\textbf{o}}),$
\\
$\mathcal{L}_{B}^{D}=D_{B}(G_{B}(\textrm{\textbf{o}},\textrm{\textbf{z}}'))-D_{B}(\textrm{\textbf{u}}),$
\\
$\mathcal{L}_{C}^{D}=D_{C}(G_{C}(\textrm{\textbf{v}},\textrm{\textbf{z}}'))-D_{C}(G_{A}(G_{B}(\textrm{\textbf{o}},\textrm{\textbf{z}}'),\textrm{\textbf{z}})),$
\\
$\mathcal{L}_{D}^{D}=D_{D}(G_{D}(G_{A}(G_{B}(\textrm{\textbf{o}},\textrm{\textbf{z}}'),\textrm{\textbf{z}}),\textrm{\textbf{z}}))-D_{D}(\textrm{\textbf{v}}),$
\\
where $D_{A}(\cdot)$, $D_{B}(\cdot)$, $D_{C}(\cdot)$ and $D_{D}(\cdot)$ are defined similarly as in \eqref{eqs:2} and \eqref{eqs:3}.
\\
With adversarial training of the proposed Dual-DualGAN, the Dual-DualGAN learns the mapping between $\textrm{\textbf{u}}\in U$ and $\textrm{\textbf{v}}\in V$.
\end{algorithmic}
\end{algorithm}

\section{Data Collection and Preprocessing}
\textbf{Participants and speech datasets.}
Participants for data collection in the study were students and academic staff from Taiyuan University of Science and Technology, all in good health, including 24 male and 26 female, aged between 20 and 40. All participants washed their hair before the experiment to ensure their scalps were clean. In addition, they were not allowed to wear any jewelry. In the experiments, the participants were asked to place their forearms and hands in a place where they feel comfortable without movements, and to relax as much as possible in order to reduce facial muscle movements and eye blinking.

The non-invasive EEG signals measuring human neural activity were collected as the participants listen to continuous speech audio with a dedicated earphone. The speech signals were taken from the TIMIT\footnote{https://catalog.ldc.upenn.edu/docs/LDC93S1/TIMIT.html} dataset which contains 6300 sentences, spoken by 630 speakers (438 male and 192 female, sampled at 16 kHz). We consider the sentences of the above dataset for training and testing.

\begin{figure}[!t]
%\begin{figure}[!htb]
%\usepackage{float}
%\begin{figure*}[H]
\centering
\includegraphics[width=3.4in]{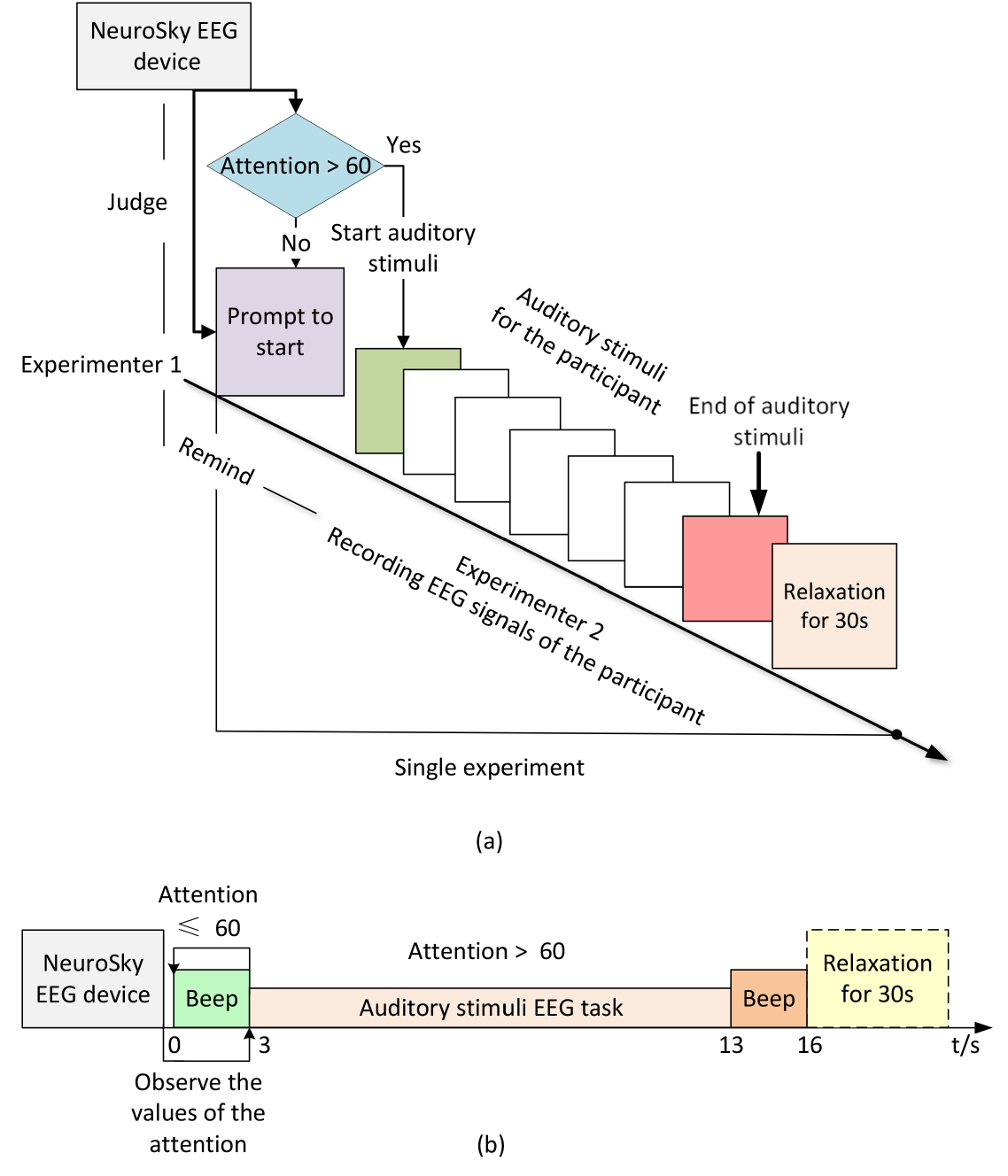}
\caption{Design of the experimental paradigm for EEG data collection (a) Experimental paradigm with supervision (b) Temporal events in each experiment.}
\label{fig:6}
\end{figure}

\textbf{Experimental paradigm with supervision.}
To improve the efficiency of auditory speech stimuli, an experimental paradigm with supervision based on the traditional Oddball experimental paradigm is proposed by considering participant's attention detected with a threshold, as illustrated in Fig. \ref{fig:6} (a). The temporal events in each experiment for data capture are shown in Fig. \ref{fig:6} (b).

\begin{figure}[!t]
%\begin{figure}[!htb]
%\usepackage{float}
%\begin{figure*}[H]
\centering
\includegraphics[width=2.3in]{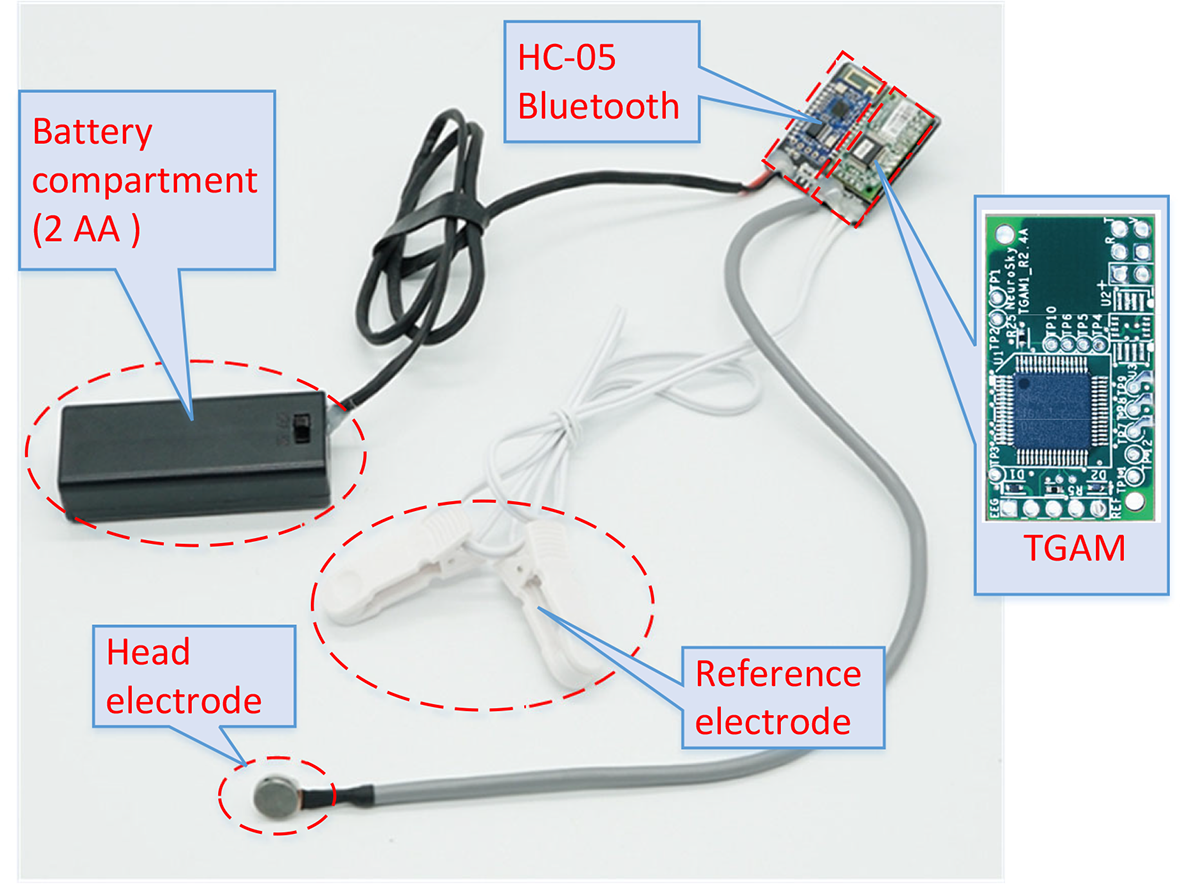}
\caption{Device for detecting participants' attention.}
\label{fig:7}
\end{figure}

To ensure the quality of the EEG signals recorded, we design a device for measuring the attention of participants in response to the stimuli played using the TGAM ($\textrm{ThinkGear}^{\textrm{TM}}$ Asic Module) produced by NeuroSky (see Fig. \ref{fig:7}). We start recording the EEG signal only when the attention is higher than a pre-defined threshold.
The attention $P$ can be described as follows based on the $\textrm{eSense}^{\textrm{TM}}$ algorithm.
\begin{equation}
P=\frac{l\beta+m\beta}{\theta},
\label{eqs:10}
\end{equation}
where $m\beta$ is middle beta waves (frequency 16-20 Hz), $l\beta$ is low beta waves (frequency 12-15 Hz), and $\theta$ is theta waves (frequency 4-7 Hz).
The attention with a value greater than a pre-defined threshold e.g. $P>60$ indicates that the participants are concentrating on the auditory stimuli, and the EEG signals can be recorded from this moment.

In Fig. \ref{fig:6}, at the beginning of an experiment, the experimenter 1 plays a beep to remind the participant concentrate to the experiment, and observes the participant's attention $P$ for 3 seconds. For $P\leq60$, the experimenter 1 repeats the above procedure. For $P>60$, the experimenter 1 starts to play a continuous speech file and reminds the experimenter 2 to record the EEG signals, and each speech file is repeated at least five times. At the end of the experiment, the experimenter 1 plays a beep to remind the experimenter 2 stopping the recording, and the participants can open their eyes, blink and relax. After relaxing for 30 seconds, they can start the next experiment.

\begin{figure}[!t]
%\begin{figure}[!htb]
%\usepackage{float}
%\begin{figure*}[H]
\centering
\includegraphics[width=3in]{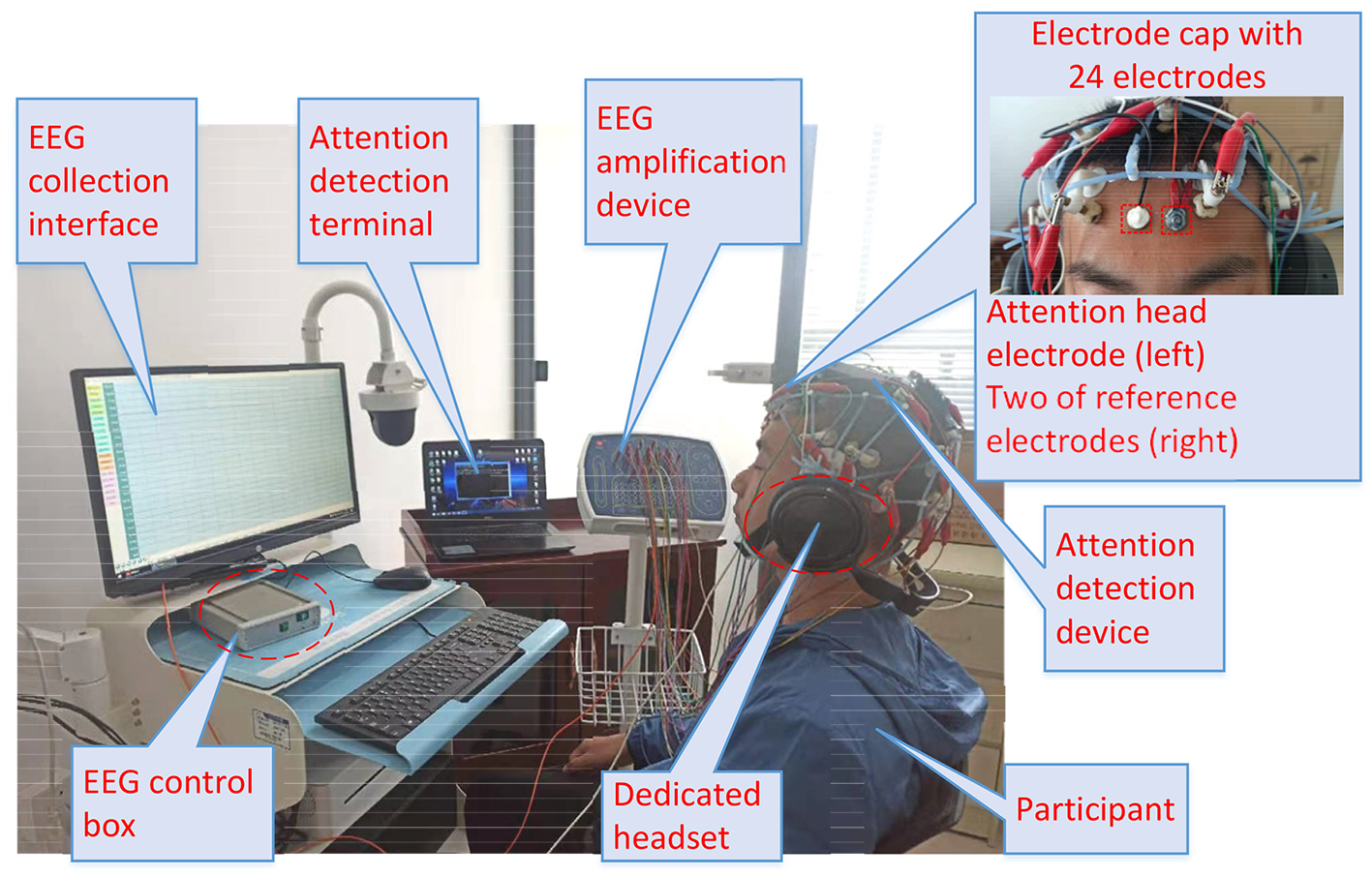}
\caption{Data collection platform for EEG signals.}
\label{fig:8}
\end{figure}

\textbf{Data collection and EEG datasets.} In the experiments, a data collection platform is set up for the non-invasive EEG neural recordings. The EEG signals are recorded from 24 electrodes placed around the scalp according to international 10-20 system by using Electroencephalogram and evoked potentiometer NCERP produced by Shanghai Nuocheng Electric Co., Ltd (NCC). By connecting the electrode cap to the physiological amplifier, analog EEG signals are collected and amplified. By using the optical fibers for transmitting the data to the EEG master control box, the amplified EEG signals are digitized at about 8 kHz and 32bit, filtered with the cut off frequencies of 1 Hz and 50 Hz, the channels with visible artifact or excessive noise are removed, and then transmitted to the computer by USB interface. The platform for the collection of EEG signals is shown in Fig. \ref{fig:8}, and the parameters of NCERP are listed in Table \ref{table_1}.

\begin{figure}[!t]
%\begin{figure}[!htb]
%\usepackage{float}
%\begin{figure*}[H]
\centering
\includegraphics[width=2in]{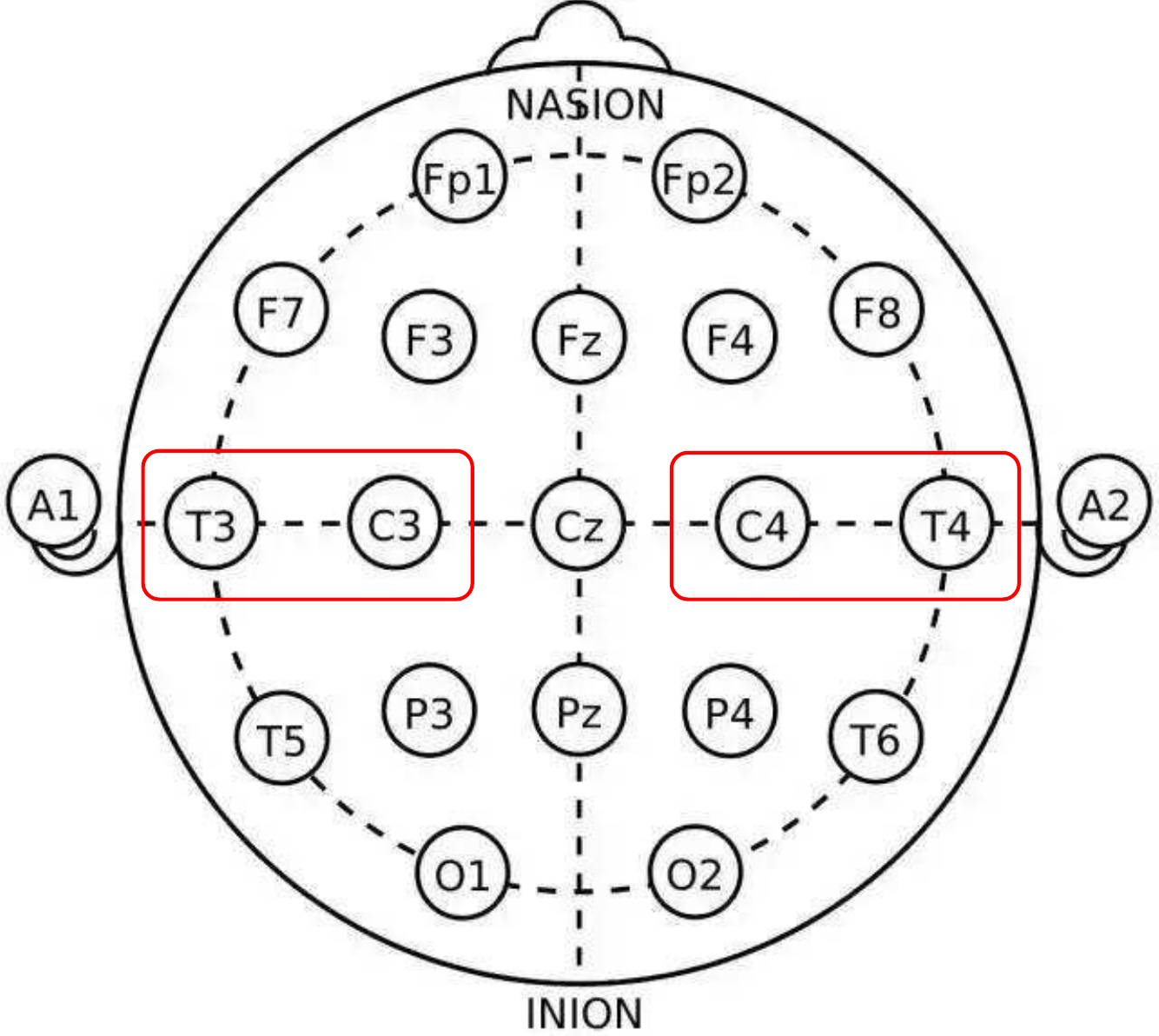}
\caption{Placement of twenty-four EEG electrodes, and four of which are selected according to the temporal lobe and highlighted with red boxes.}
\label{fig:9}
\end{figure}

\begin{table}[htbp]
    \renewcommand{\arraystretch}{1.3}
    \renewcommand\tabcolsep{1.8pt}
	\caption{The parameters of NCERP.}
	\centering
	\label{table_1}
		\begin{tabular}{cc}
			\hline\hline \\[-3mm]
            Parameter & Value \\[0.5ex]\hline
            Calibration voltage & 100 $\mu$V, and error is not more than $\pm$ 5 $\%$\\
            Sensitivity & 5 $\mu$V/cm, and error is not more than $\pm$ 5 $\%$\\
            Time constants & 0.1 s, 0.2 s, 0.3 s, and error is $\leq\pm$ 10 $\%$\\
            Noise level & $<$ 0.3 $\mu$V (RMS)\\
            Common mode rejection ratio & $\geq$ 110 dB\\
            Amplitude frequency & \multirow{2}*{1 Hz $\sim$ 60 Hz, and error + 3 $\%\sim$ - 15 $\%$}\\
            characteristic & \\
            \multirow{2}*{Polarization resistance voltage} & $\pm$ 300 mV DC polarization voltage,\\
            & sensitive change $\pm$ 5 $\%$\\
            Input impedance & $\geq$ 10 M$\Omega$
            \\[0.5ex]		
			\hline\hline
		\end{tabular}
\end{table}

We choose four-channel EEG signals according to the EEG electrode position of the temporal lobe (see Fig. \ref{fig:9} with red boxes). The selected EEG signals are normalized and transformed into a multi-dimensional matrix to build a new EEG dataset for the participants with good attention, which will be the input of the proposed Dual-DualGAN.

\section{Experimental Setup}
In our Dual-DualGAN, the generators are the U-shaped net as in
\cite{zhou2018unet++} configured with equal number of down-sampling (pooling) and up-sampling layers. The down-sampling (pooling) layers are constructed by eight convolution layers with the kernel of (3, 3), each neuron with a LeakyReLU activation function, the step of convolution is 1, and the step of pooling is 2. The up-sampling layers are constructed by eight deconvolution layers with the kernel of (3, 3), each neuron with a LeakyReLU activation function, and the step of convolution is 1. The skip connections between mirrored down-sampling and up-sampling layers are used to enable the low-level information to be shared between the input and output and to avoid loss of information. The discriminators are the Markovian Patch-GAN as in
\cite{li2016precomputed}, which are constructed by five convolution layers with the kernel of (3, 3), and has no constraints over the size of the input signal. The number of critic iterations per generator iteration $N$ can be set to 5, $\lambda_{U}$, $\lambda_{U}$ and $\lambda_{O}$ are all set to 500, an initial learning rate is set at 0.0002, and the batch size $M$ is assigned with 1.

\textbf{Baseline methods.} We compare several versions of our Dual-DualGAN algorithm: M1 (removing two convolution layers of the generative network), M2 (increasing two convolution layers of the generative network), M3 (removing two convolution layers of the discriminative network), M4 (increasing two convolution layers of the discriminative network), M5 (setting the step of convolution to 2), M6 (setting the convolution layers with the kernel of (5, 5)), M7 (changing activation function to ReLU), Dual-DualGAN (full version of our Dual-DualGAN algorithm) with the state-of-the-art algorithms: Deep Neural Network (DNN)-based encoder-decoder algorithm
\cite{akbari2019towards},
bidirectional Long and Short Term Memory Network (bLSTM)-based encoder-decoder algorithm
\cite{anumanchipalli2019speech},
Recurrent Neural Network (RNN)-based encoder-decoder algorithm
\cite{krishna2020speech},
and DualGAN \footnote{https://github.com/duxingren14/DualGAN}
\cite{yi2017dualgan}.

The input to the DNN-based, bLSTM-based encoder-decoder algorithms as in
\cite{akbari2019towards},
\cite{anumanchipalli2019speech} is intracranial ECoG signal, whereas the input to our Dual-DualGAN is non-invasive EEG signal. The RNN-based encoder-decoder algorithm in
\cite{krishna2020speech} does not aim to reconstruct the speech signal, but predict acoustical feature of MFCC. Thus these codes cannot be used directly. Based on an encoder-decoder network \footnote{https://github.com/tensorflow/nmt}, DNN-based, bLSTM-based, and RNN-based encoder-decoder algorithms are coded by learning from the methods as in
\cite{akbari2019towards},
\cite{anumanchipalli2019speech},
\cite{krishna2020speech}.

The DNN-based encoder-decoder algorithm consists of two modules, namely, feature extraction and feature summation networks. The feature extraction network for auditory spectrogram reconstruction is a convolutional neural network (CNN) constructed by four convolution layers with the kernel of (3, 3), each neuron with a LeakyReLU activation function and a dropout of 0.3. The feature summation network is a two-layer fully connected neural network (FCN). Each FCN is constructed by three fully connected layer of 256, each neuron with a LeakyReLU activation function and a dropout of 0.3.

The bLSTM-based encoder-decoder algorithm is a stacked encoder-decoder network.
The encoder is implemented as two feedforward layers followed by two-layer bLSTM. The
decoder is implemented as three-layer bLSTM. Each bLSTM is constructed by a forward LSTM and a backward LSTM, and each LSTM cell has 100 hidden units. Training of the models is stopped when the validation loss no longer decreases with an initial learning rate set at 0.001, a dropout of 0.4, and the batch size set at 25.

The RNN-based encoder-decoder algorithm consists of two modules: encoder (acoustic feature extraction) and decoder (reconstruction).
The encoder and the decoder are constructed by two layers of gated recurrent unit (GRU) with 256 hidden units in the first layer and 128 hidden units in the second layer. The final GRU layer is connected to a fully connected layer of 13 hidden units at each step.
The model is trained for 250 epochs with an initial learning rate set at 0.01, a dropout rate of 0.2, and the batch size set at 100.

All the algorithms are trained from scratch by using cross validation, by randomly picking 80 $\%$ data of the EEG dataset for training, the remaining 20 $\%$ data for testing. The facilities used to perform the experiments include Intel I9-10900X 13.7 GHz CPU, 2*NVIDIA RTX 8000 Graphics Card and 6*32 GB memory.

For performance evaluation, we use the accuracy rate
\cite{story1986accuracy}, Pearson correlation coefficient (PCC)
\cite{benesty2009pearson} and Mel-cepstral distortion (MCD)
\cite{Kubichek1993MCD} as the performance metrics.

The accuracy rate is the proportion of correct predictions among the total number of cases examined.
\begin{equation}
Accuracy=\frac{T}{T+F},
\label{eqs:11}
\end{equation}
where $T$ means the correct predictions and $F$ means the false predictions.

The PCC is a measure of linear correlation between the original and the synthesized speech signal, defined as
\begin{equation}
PCC=\frac{\textrm{cov}(\textbf{v}\textbf{v}^{'})}{\sigma_{\textbf{v}}\sigma_{\textbf{v}^{'}}},
\label{eqs:12}
\end{equation}
where $\textrm{cov}(\textbf{v}\textbf{v}^{'})$ is the covariance of the original speech signal $\textbf{v}$ and the synthesized speech signal $\textbf{v}^{'}$, and $\sigma_{\textbf{v}}$ and $\sigma_{\textbf{v}^{'}}$ are the standard deviation of $\textbf{v}$ and $\textbf{v}^{'}$, respectively.

The metric $MCD(k)$ evaluates objective speech quality, defined as
\begin{equation}
MCD(k)=\frac{10}{\ln 10}\frac{1}{T}\sum_{i=0}^{T-1}\sqrt{\sum_{k=1}^{K}(mc(i,k)-mc'(i,k))^{2}},
\label{eqs:13}
\end{equation}
where $mc(i,k)$ and $mc'(i,k)$ are the $i$-th mel-cepstral coefficient of the $k$-th frame of the original and the synthesized speech signal, respectively.

\section{Results}
In this section, we carry out experiments to demonstrate the performance of the proposed Dual-DualGAN, and how the algorithm is affected for solving the ET-CAS problem.

To evaluate the performance of our Dual-DualGAN, we conduct three listening tasks that involve word-level, short-sentence-level (no more than six words) and long-sentence-level (more than six words) transcription, respectively. The word-level speech signals are separated from the sentence-level speech signals.

In Table \ref{table_2}, we compare several versions of our Dual-DualGAN algorithm with the state-of-the-art algorithms. The results show that the proposed Dual-DualGAN has better performance in average accuracy rate, PCC and MCD than the RNN-based, bLSTM-based, DNN-based encoder-decoder algorithms, and DualGAN. The accuracy rates and PCCs of the RNN-based, bLSTM-based, DNN-based encoder-decoder algorithms, and DualGAN are less than 74.9$\%$ and 0.78, and the MCDs of the RNN-based, bLSTM-based, DNN-based encoder-decoder algorithms, and DualGAN are more than 4.1 dB.
\begin{table}[htbp]
    \renewcommand{\arraystretch}{1.3}
    \renewcommand\tabcolsep{4pt}
	\caption{The performance and time complexity of the Dual-DualGAN as compared with
state-of-the-art algorithms.}
	\centering
	\label{table_2}
		\begin{tabular}{ccccc}
			\hline\hline \\[-3mm]
            \multicolumn{2}{c}{\multirow{2}*{Algorithm}} & Accuracy rate & PCC & MCD \\
            & & ($\%$) &  & (dB)\\[0.5ex]\hline
            \multirow{3}*{\shortstack{Encoder-decoder\\framework}}& DNN & 74.87 & 0.771 & 4.154\\
            & bLSTM & 70.04 & 0.729 & 4.437\\
            & RNN & 71.10 & 0.747 & 4.352\\
            \multicolumn{2}{c}{DualGAN} & 56.82 & 0.624 & 5.015\\[0.5ex]	
			\hline
            \multicolumn{2}{c}{M1} & 74.24 & 0.792 & 3.947\\
            \multicolumn{2}{c}{M2} & 77.92 & 0.829 & 3.808\\
            \multicolumn{2}{c}{M3} & 73.86 & 0.788 & 3.971\\
            \multicolumn{2}{c}{M4} & 78.18 & 0.834 & 3.798\\
            \multicolumn{2}{c}{M5} & 77.03 & 0.821 & 3.815\\
            \multicolumn{2}{c}{M6} & 77.48 & 0.828 & 3.811\\
            \multicolumn{2}{c}{M7} & 75.83 & 0.796 & 3.941\\
            \multicolumn{2}{c}{Dual-DualGAN} & \textbf{78.53} & \textbf{0.838} & \textbf{3.793}\\[0.5ex]
			\hline\hline
		\end{tabular}
\end{table}

\begin{figure}[!htb]
%\begin{figure}[!htb]
%\usepackage{float}
%\begin{figure*}[H]
\centering
\includegraphics[width=3.3in]{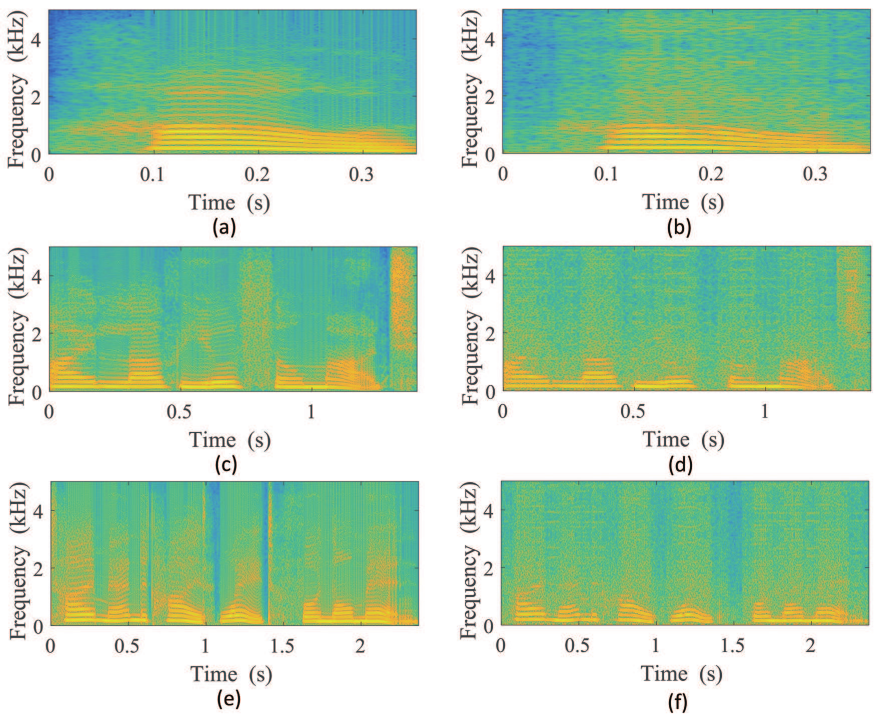}
\caption{Audio spectrograms of the synthesized speech and the original speech signals (a) Original spectrogram of a word speech signal (b) Synthesized spectrogram of the word speech signal (c) Original spectrogram of a short sentence speech signal (d) Synthesized spectrogram of the short sentence speech signal (e) Original spectrogram of a long sentence speech signal (f) Synthesized spectrogram of the long sentence speech signal.}
\label{fig:10}
\end{figure}

Fig. \ref{fig:10} shows the audio spectrograms from the original word, short sentence and long sentence speech signals, respectively, and those decoded from human neural activity. All the synthesized spectrograms retain salient energy patterns that are present in the original spectrograms.

As listed in Table \ref{table_3}, we find that listeners are more successful at word-level transcription, and the values of the accuracy rates can reach 100 $\%$. Although the synthesized spectrograms of the short sentence and long sentence speech signals retain salient energy patterns of the original spectrograms, the values of the accuracy rates have dropped to less than 80 $\%$.
\begin{table}[htbp]
    \renewcommand{\arraystretch}{1.3}
    \renewcommand\tabcolsep{1.8pt}
	\caption{Listener transcriptions of neurally synthesized speech for different words and sentences.}
	\centering
	\label{table_3}
		\begin{tabular}{ccl}
			\hline\hline \\[-3mm]
            \multirow{2}*{Type} & Accuracy & Original words or sentences and \\
            & rate & transcriptions of synthesized speech \\[0.5ex]\hline
            \multirow{2}*{Word} & \multirow{2}*{100$\%$} & O: weekend \\
            & & T: weekend \\[0.5ex]\hline
            \multirow{4}*{Short sentence} & \multirow{2}*{75$\%$} & O: Happy birthday to you.\\
            & & T: Have birthday to you. \\
            & \multirow{2}*{50$\%$} & O: Bob likes to play with dog. \\
            & & T: Bob bikes do play with hot. \\[0.5ex]\hline
            \multirow{8}*{Long sentence} & \multirow{4}*{60$\%$} & O: What makes the desert beautiful is that \\
            & & somewhere it hides a well. \\
            & & T: What made the desert beautiful was that \\
            & & flower it his a dog. \\
            & \multirow{4}*{50$\%$} & O: The stars are beautiful, because of a \\
            & & flower that cannot be seen. \\
            & & T: The was well beautiful, beautiful of a \\
            & & somewhere what cannot a seen. \\[0.5ex]		
			\hline\hline
		\end{tabular}
\end{table}

\begin{figure*}[!htb]
%\begin{figure}[!htb]
%\usepackage{float}
%\begin{figure*}[H]
\centering
\includegraphics[width=5.5in]{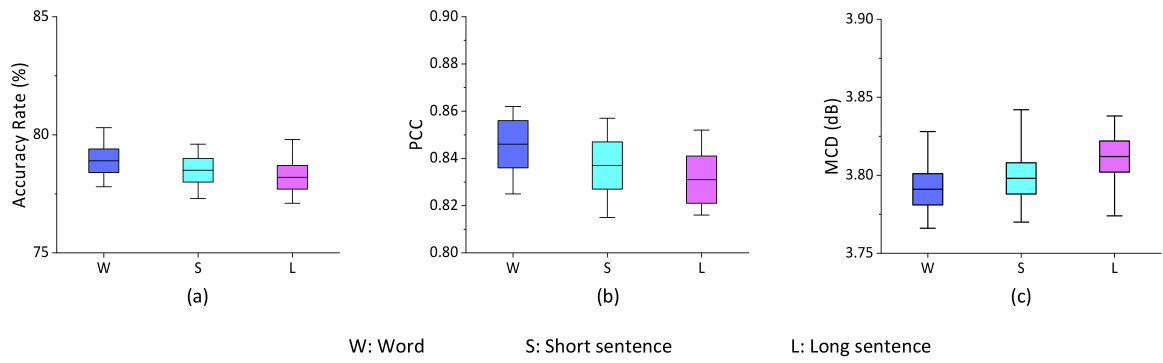}
\caption{Performance evaluations of the neurally synthesized speech for the word-level, short-sentence-level and long-sentence-level transcription (a) Average accuracy rate of the synthesized speech signals (b) Average PCC between the synthesized speech signals and the original speech signals (c) Average MCD of the synthesized speech signals in comparison with the original speech signals.}
\label{fig:11}
\end{figure*}

In Fig. \ref{fig:11} (a), we compare the average accuracy rates of the synthesized speech signals for word-level, short-sentence-level and long-sentence-level transcription. The average accuracy rate of the word-level transcription is higher than those of the short-sentence-level and long-sentence-level transcription, and all the values of the accuracy rate are around 78.5 $\%$. The average PCC between the synthesized speech signals and the original speech signals are shown in Fig. \ref{fig:11} (b). The PCCs of the word-level, short-sentence-level and long-sentence-level transcription are relatively high and the values are above 0.83. We also compare the average MCD of the synthesized speech signals (see Fig. \ref{fig:11} (c)). The MCDs of the synthesized speech signals for word-level, short-sentence-level and long-sentence-level transcription are relatively small, and the values of MCDs are about 3.9. This demonstrates the efficiency of the proposed Dual-DualGAN in decoding speech from human neural activity to synthesized speech.

\begin{figure}[!t]
%\begin{figure}[!htb]
%\usepackage{float}
%\begin{figure*}[H]
\centering
\includegraphics[width=3in]{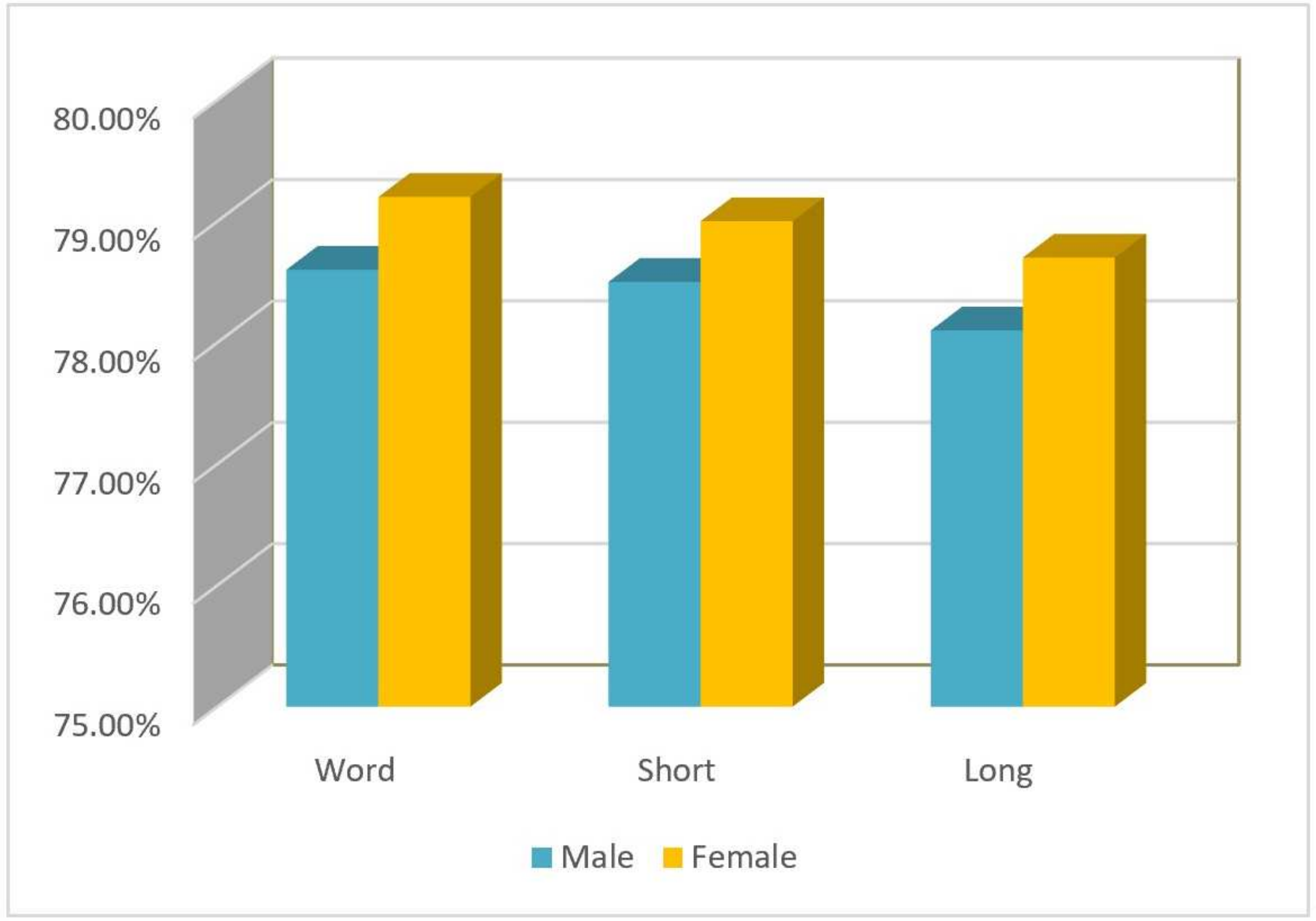}
\caption{Average accuracy rates of the synthesized speech signals for word-level, short-sentence-level and long-sentence-level transcription by gender.}
\label{fig:12}
\end{figure}

The gender effects are considered on listener transcriptions of the neurally synthesized speech. The data of 20 male and 20 female are randomly selected from the EEG datasets. The average accuracy rates of the synthesized speech signals for the word-level, short-sentence-level and long-sentence-level transcription by gender are illustrated in Fig. \ref{fig:12}. The average accuracy rates of male for word-level, short-sentence-level and long-sentence-level transcription are 78.5 $\%$, 78.3 $\%$, and
77.9 $\%$, and the average accuracy rates of female are 79.1 $\%$, 78.9 $\%$, and
78.6 $\%$. It shows the efficiency and adaptability of the proposed Dual-DualGAN, regardless of gender.

\begin{figure}[!t]
%\begin{figure}[!htb]
%\usepackage{float}
%\begin{figure*}[H]
\centering
\includegraphics[width=3in]{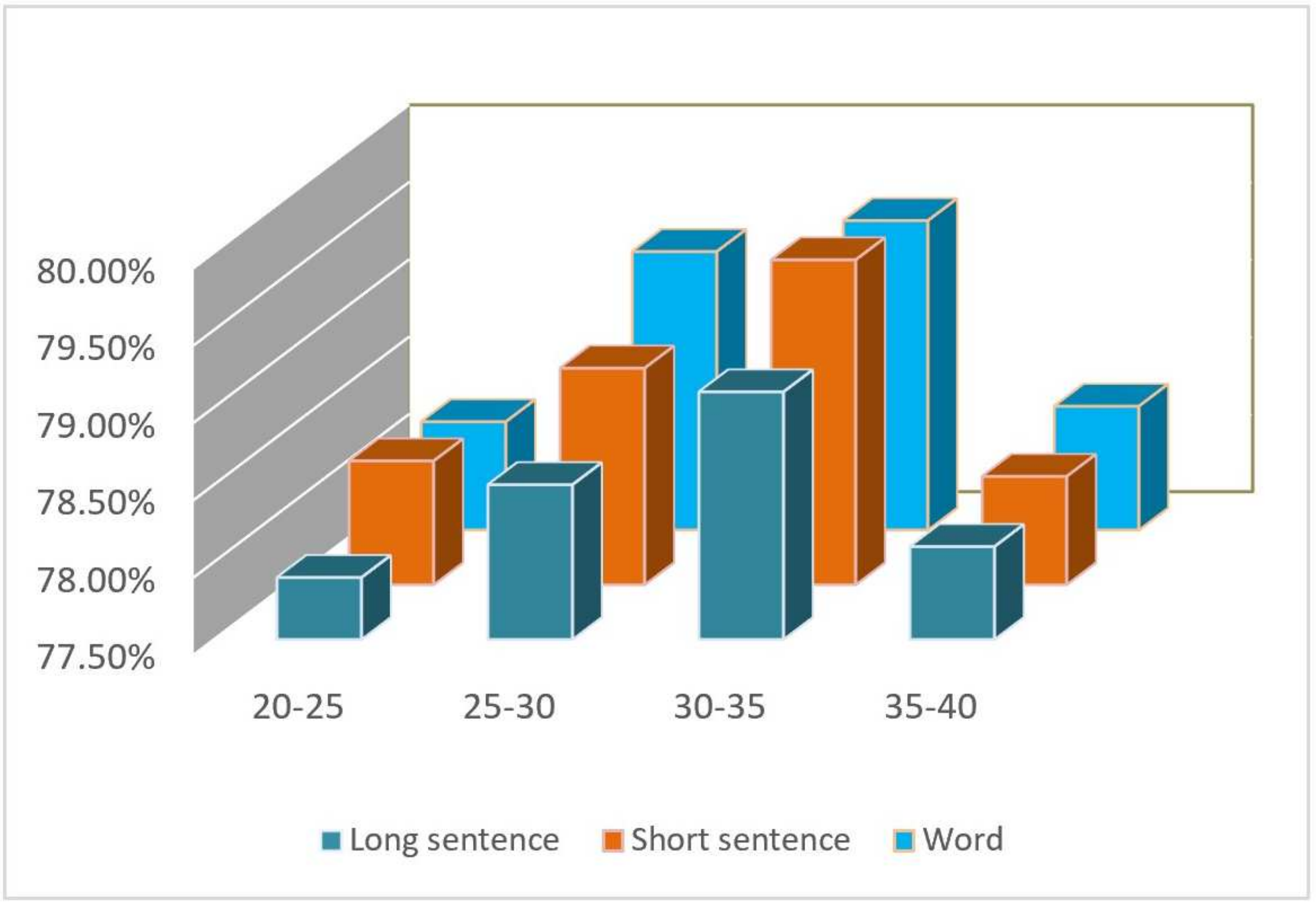}
\caption{Average accuracy rates of the synthesized speech signals for word-level, short-sentence-level and long-sentence-level transcription by age.}
\label{fig:13}
\end{figure}

The age effects are also considered on listener transcriptions of the neurally synthesized speech. The data of four age groups (20-25, 25-30, 30-35, and 35-40 years old) of the participants are randomly picked from the EEG datasets. The average accuracy rates of the synthesized speech signals for word-level, short-sentence-level and long-sentence-level transcription for these age groups are illustrated in Fig. \ref{fig:13}. The average accuracy rates for word-level, short-sentence-level and long-sentence-level transcription of the participants in age between 25 to 30 and 30 to 35 years old are mostly above 78.5 $\%$. The average accuracy rates of the participants in age between 20 to 25 and 35 to 40 years old are slightly low, and the values are about 78 $\%$. The results demonstrate the proposed Dual-DualGAN can translate an EEG to a speech signal with a good generalization ability for different age groups.

\section{Conclusion}
We have presented a new method for the problem of end-to-end translation from human neural activity to speech (ET-CAS). Our contributions to this challenging problem are as follows:

\textbf{Model.} We have formulated an end-to-end model for the ET-CAS problem, i.e. translating human neural activity to speech directly.

\textbf{Datasets.} We developed a new EEG dataset where the attention of the participants is detected and used to guide the collection of the EEG signals in each experiment.

\textbf{Network.} We proposed a dual-dual generative adversarial network (Dual-DualGAN) to address the ET-CAS problem. In this system, two DualGAN are created and trained simultaneously, where a transition domain is introduced into the DualGAN to bridge the two DualGAN. The EEG and speech signals are cascaded proportionally to generate the transition signals i.e. constructing shared labels for EEG and speech signals without mapping their features.

Numerical experiments show that the proposed ET-CAS algorithm performs well in translating human neural activity to speech. In the future, it is interesting to investigate how to incorporate acoustic and emotional features into the ET-CAS model and algorithm, which may improve the performance of the system in decoding speech signals that consist of sentences with repetitive words.

%\newpage
% if have a single appendix:
%\appendix[Proof of the Zonklar Equations]
% or
%\appendix  % for no appendix heading
% do not use \section anymore after \appendix, only \section*
% is possibly needed

% use appendices with more than one appendix
% then use \section to start each appendix
% you must declare a \section before using any
% \subsection or using \label (\appendices by itself
% starts a section numbered zero.)
%

% Can use something like this to put references on a page
% by themselves when using endfloat and the captionsoff option.
%\ifCLASSOPTIONcaptionsoff
%  \newpage
%\fi

% trigger a \newpage just before the given reference
% number - used to balance the columns on the last page
% adjust value as needed - may need to be readjusted if
% the document is modified later
%\IEEEtriggeratref{8}
% The "triggered" command can be changed if desired:
%\IEEEtriggercmd{\enlargethispage{-5in}}

% references section

% can use a bibliography generated by BibTeX as a .bbl file
% BibTeX documentation can be easily obtained at:
% http://mirror.ctan.org/biblio/bibtex/contrib/doc/
% The IEEEtran BibTeX style support page is at:
% http://www.michaelshell.org/tex/ieeetran/bibtex/
\bibliographystyle{IEEEtran}
% argument is your BibTeX string definitions and bibliography database(s)
%\bibliography{IEEEabrv,../bib/paper}
%
% <OR> manually copy in the resultant .bbl file
% set second argument of \begin to the number of references
% (used to reserve space for the reference number labels box)
%\begin{thebibliography}{1}

%\bibitem{IEEEhowto:kopka}
%J.~Kulmer and P.~Mowlaee, ``Phase estimation in single channelfig:2 speech enhancement using phase decomposition," \emph{IEEE Signal Process. Lett.}, vol. 22, no. 5, pp. 598--602, 2015.

\bibliography{bibli}

\end{document}